\documentclass[useAMS,usenatbib,referee]{biom}
\def\bSig\mathbf{\tau}

\usepackage[figuresright]{rotating}
\usepackage{multicol, multirow}
\usepackage{amsmath,amssymb}
\usepackage{mathrsfs}
\usepackage{natbib}
\usepackage{hyperref}
\usepackage{bigints}
\usepackage{xcolor}

\title[Copas-Jackson-type bounds]{Copas-Jackson-type bounds for publication bias over a general class of selection models}

\author{Taojun Hu\\
	 Department of Biomedical Statistics, Graduate School of Medicine, Osaka University, 565-0871, Osaka, Japan\\
 Department of Biostatistics, School of Public Health, Peking University, 100191, Beijing, China
	 \and 
	 Yi Zhou\\
	 Beijing International Center for Mathematical Research, Peking University, 100871, Beijing, China\\
     Department of Biomedical Statistics, Graduate School of Medicine, Osaka University, 565-0871, Osaka, Japan
     \and
     Xiao-Hua Zhou\\
     Beijing International Center for Mathematical Research, Peking University, 100871, Beijing, China\\
     Department of Biostatistics, School of Public Health, Peking University, 100191, Beijing, China\\
 \and
 Satoshi Hattori$^{*}$\email{hattoris@biostat.med.osaka-u.ac.jp}\\
	 Department of Biomedical Statistics, Graduate School of Medicine, Osaka University, 565-0871, Osaka, Japan\\
 Integrated Frontier Research for Medical Science Division, \\Institute for Open and Transdisciplinary Research Initiatives (OTRI), Osaka University, 565-0871, Osaka, Japan\\
 }

\begin{document}





\label{firstpage}


\begin{abstract}
Publication bias (PB) is one of the most vital threats to the accuracy of meta-analysis. Adjustment or sensitivity analysis based on selection models, which describe the probability of a study being published, provide a more objective evaluation of PB than widely-used simple graphical methods such as the trim-and-fill method. Most existing methods rely on parametric selection models. The Copas-Jackson bound (C-J bound) provides a worst-case bound of an analytical form over a nonparametric class of selection models, which would provide more robust conclusions than parametric sensitivity analysis. The nonparametric class of the selection models in the C-J bound is restrictive and only covers parametric selection models monotonic to the standard errors of outcomes. The novelty of this paper is to develop a method that constructs worst-case bounds over a general class of selection models weakening the assumption in the C-J bound. We propose an efficient numerical method to obtain an approximate worst-case bound via tractable nonlinear programming with linear constraints. We substantiate the effectiveness of the proposed bound with extensive simulation studies and show its applicability with two real-world meta-analyses.
\end{abstract}

%



\maketitle

\section{Introduction}
\label{sec:intro}

Systematic reviews collect the relevant studies on a specific research question. A single individual study may yield unilateral results based on potentially biased sampling from the population. Thus, systematic reviews lend themselves to drawing comprehensive conclusions. Meta-analysis is used to synthesize results from each study and form integrated findings quantitatively. In this paper, we are interested in estimating the effect size, such as the log odds ratio (lnOR), of an experimental treatment based on summary statistics reported in scientific journals. Suppose that each study reports its estimate of the lnOR and the standard error. Such data can be extracted from multiple literatures. Then, meta-analysis synthesizes these results and offers an overall estimate of the lnOR. However, meta-analysis with only published studies might suffer from publication bias (PB), which refers to the bias due to selective publication in scientific journals. Scientific journals tend to prefer studies with expected and significant results; those with unexpected and insignificant results are less likely to be published. A large number of literatures recognize the impact of PB in meta-analysis and try to address PB with different methods, which can be majorly categorized into graphical methods and quantitative methods based on selection models. A good review of methods addressing PB can be found in~\citet{jin2015statistical}. 

Graphical methods, such as the funnel-plot and the trim-and-fill method, address PB relying on the asymmetry of the funnel-plot~\citep{duval2000trim, duval2000nonparametric, light1984summing}. Some tests to detect funnel-plot asymmetry are available such as Begg's rank test~\citep{begg1994operating} and Egger's regression tests~\citep{egger1997bias}. These methods based on funnel-plot asymmetry are simple and are widely used in practice. However, selective publication is not the only cause of asymmetry of the funnel plot, and these methods may provide very subjective conclusions. Alternatively, methods based on selection models are available. They quantitatively model the underlying selective publication mechanism, thus leading to less subjective results compared with graphical methods~\citep{carpenter2009empirical}. The quantitative methods can be broadly categorized into two types: the weighted methods~\citep{dear1992approach, iyengar1988selection, preston2004adjusting, rosenthal1988selection} and the sensitivity analysis methods~\citep{copas2001sensitivity, copas2013likelihood, hattori2018sensitivity, hu2024sensitivity, zhou2023likelihood}. Most of the weighted methods require large sample size to obtain the robust estimation, which limits their use in meta-analysis where only a handful of primary studies are available. On the other hand, the sensitivity analysis methods allow sensitivity parameters and can handle small-scale meta-analysis, thus more appealing. In the present paper, we focus on the sensitivity analysis methods while a brief summarization of the weighted methods can be referred to~\citep{jin2015statistical}.

Among the sensitivity analysis methods, \citet{copas1999works} and~\citet{copas2000meta, copas2001sensitivity} assumed that publication of a study was determined by a latent Gaussian random variable, and the random variable was related to the standard deviation and the effect size of each study. The model borrows ideas from the Heckman statistics which were first applied in economics~\citep{heckman1976common, heckman1979sample}. We call it the Copas-Heckman selection model in this paper. It has been found that the parameters in the Copas-Heckman selection model are difficult to estimate simultaneously~\citep{carpenter2009empirical}. As an alternative, \citet{copas2013likelihood} proposed the sensitivity analysis method based on a simpler method which is easier to implement and interpret. This method assumed the selection probability of a study was a monotone function of the $t$-statistics, or, equivalently, the $p$-values, in the individual studies. There had been other previous methods relying on the $t$-statistics type selection models~\citep{iyengar1988selection,preston2004adjusting} but \citet{copas2013likelihood} is the first to incorporate them into sensitivity analysis. Their sensitivity analysis told us how the estimated effect size would change with different proportions of published studies among all studies. Due to its interpretability, it has been extended to various types of meta-analysis, including the meta-analysis of diagnostic studies, the network meta-analysis, and the meta-analysis of sparse data~\citep{zhou2023likelihood, hu2024sensitivity, huang2024publication}. 

All the above sensitivity analysis methods, as well as the weighted methods in~\citep{jin2015statistical}, are based on some specific parametric assumptions on the selection models. Different choices of the parametric selection models would give varied results, and interpretations of the results may lead to various conclusions. In addition, it is hard to select a suitable parametric selection model to describe the true selection mechanism with limited information from a meta-analysis. To this end, methods allowing flexible selection models are more appealing. By allowing flexible choices of selection models, one can find out the potential maximal (worst-case) bias on the estimation of the overall effect size in a meta-analysis. \citet{copas2004bound} is the first study to evaluate the lower and upper bound of PB over a class of nonparametric selection models under some given numbers of unpublished studies. Their bound describes the worst-case PB under a variety of selection models. It assumes the publication probability is a non-decreasing function to the variances of the effect sizes, which indicates the publication probability of a study with larger sample sizes is larger than those with smaller sample sizes. The bounds obtained under this assumption is called the Copas-Jackson bound (C-J bound). \citet{zhou2024nonparametric} realized the C-J bound via a simulation-based algorithm. However, the examination of the connection between the existing parametric selection models and the assumptions of the C-J bound is lacking~\citep{copas2004bound}. It should be verified whether the assumptions of~\citep{copas2004bound} are satisfied by the various widely used selection models. If not, it is necessary to develop a new sensitivity method deriving the bounds of PB under a novel assumption that can suit most typical selection models.

In this paper, we first examine whether the assumptions by~\citet{copas2004bound} are satisfied by some typical selection models. We found that the Copas-Heckman selection model satisfies the assumption of the C-J bound. However, at least some $t$-statistics type selection models do not satisfy the assumption. Thus, the Copas-Jackson bounds may not cover a sufficiently large class of selection models. To handle this issue, we propose an alternative assumption to describe a wider class of selection models. Then, we propose a simulation-based sensitivity method to derive the bounds under the general class of selection models, which cover the Copas-Heckman and $t$-statistics type selection models. We illustrate the effectiveness and reliability of our method by extensive simulation studies and two real-world meta-analyses. This paper is organized as follows: in Section~\ref{sec:pre}, we introduce some preliminaries about the models for meta-analysis, widely used selection models, and the sensitivity analysis by~\citet{copas2004bound}; in Section~\ref{sec:connection}, we analyze if the assumption of C-J bound is satisfied by some of these selection models. We propose our relaxed assumption for selection models and the simulation-based method to obtain the bounds of PB under this assumption in Section~\ref{sec:bound}. We further substantiate the effectiveness of the extended bound and compare it with the C-J bound in Section~\ref{sec:simu} through a simulation study. We illustrate the applicability of our proposed methods with two real-world examples in Section~\ref{sec:ex}. Section~\ref{sec:disc} discusses and concludes this paper.

\section{Preliminaries: selection models and the Copas-Jackson nonparametric worst-case bounds for publication bias}
\label{sec:pre}

\subsection{Random-effects meta-analysis for intervention studies}
\label{sec:re}

Suppose that we are conducting a meta-analysis of $N$ published studies. The meta-analysis aims to compare the effect of two treatments: the experimental treatment and the control treatment. We are interested in evaluating the treatment effect measured by the effect size, such as the lnOR. We denote the empirical estimates of the treatment effect in the $i~(i=1, \cdots, N)$-th study as $y_i$, and the corresponding standard error as $s_i$. Following the convention in meta-analysis, $s_i$ is regarded as known. Let $\mu_i$ be the true value of the effect size in the $i$-th study. The random-effects model assumes $y_i\sim N(\mu_i, s_i^2)$ conditional on $\mu_i$, and $\mu_i \sim N(\mu, \tau^2)$, where $\mu$ is the overall effect size and $\tau^2$ is the between-study variance. They lead to the marginal model $y_i\sim N(\mu, \tau^2+s_i^2)$. The unknown parameters can be estimated through the maximized likelihood estimation (ML), the restricted maximum likelihood estimation (REML)~\citep{rukhin2000restricted}, and the moment-based estimation~\citep{dersimonian1986meta}.

\subsection{Selection models}
\label{sec:sm}

As aforementioned, the selection models play a central role in both the weighted PB-adjusted method and the sensitivity analysis. The most widely used selection models include the Copas-Heckman selection models and the $t$-statistics type selection models. The Copas-Heckman selection models assume a latent random variable $Z_i$ which determines if the study is published~\citep{copas1999works, copas2000meta, copas2001sensitivity}. To express the Copas-Heckman selection model, we reformulate the random-effects model as follows, 
\begin{equation*}
 y_i = \mu_i + s_i\epsilon_i, ~~ \epsilon_i \sim N(0, 1).
\end{equation*}
$Z_i$ is modeled as follows, 
\begin{equation*}
 Z_i=\gamma_0+\frac{\gamma_1}{s_i}+\delta_i, \delta_i \sim N(0, 1), \operatorname{corr}\left(\delta_i, \epsilon_i\right)=\rho, 
\end{equation*}
where $(\gamma_0, \gamma_1)$ are parameters to control the publication process irrespective of the empirical effect size in each published study. The study is regarded as published if and only if $Z_i>0$. Parameter $\gamma_1$ is often assumed to be positive as studies with smaller variances or, equivalently, larger sample sizes are more likely published; $\rho$ describes the influence of $y_i$ on the publication process. The publication probability conditional on observed $y_i$ and $s_i$, denoted by $p(y_i, s_i)$, is called the selection model. In the Copas-Heckman selection model, the selection model is given by
\begin{equation}
 p\left(y_i, s_i\right)=P\left(Z_i>0 \mid y_i, s_i\right)=\Phi\left\{\frac{\gamma_0+\frac{\gamma_1}{s_i}+\frac{\rho s_i\left(y_i-\mu\right)}{\tau^2+s_i^2}}{\sqrt{1-\frac{\rho^2 s_i^2}{\tau^2+s_i^2}}}\right\}, 
 \label{eq.heckman}
\end{equation}
where $\Phi(\cdot)$ is the cumulative distribution function of the standard normal distribution, also known as the probit function. When $\rho=0$, $Z_i$ is independent of $y_i$, suggesting the absence of PB.

The $t$-statistics type selection model describes the publication probability as a monotone function of the $t$-statistics. It has been recognized that studies with significant findings are more likely to be published. The significant findings are often associated with the $p$-values or its corresponding $t$-statistics. Thus, the monotonicity assumption between the publication probability and reported $t$-statistics is reasonable. \citet{preston2004adjusting} utilized some one-parameter selection models including the logistic function
\begin{equation*} 
 p\left(y_i, s_i\right)= H(t_i) = \frac{2 \exp \left[-\beta\left\{1-\Phi\left(t_i\right)\right\}\right]}{1+\exp \left[-\beta\left\{1-\Phi\left(t_i\right)\right\}\right]}, 
\end{equation*}
where $t_i=y_i/s_i$ is the $t$-statistics for the $i$-th individual study, and the modified one-parameter logistic function, 
\begin{equation*}
 p\left(y_i, s_i\right)= H(t_i, s_i)=\frac{2 \exp \left[-\beta s_i\left\{1-\Phi\left(t_i\right\}\right)\right]}{1+\exp \left[-\beta s_i\left\{1-\Phi\left(t_i\right\}\right)\right]}.
\end{equation*}
For abbreviation, we denote them by the \textit{1-logit} and \textit{mlogit} selection models, respectively. $\beta$ is the parameter that determines how the variation of $t_i$ will affect the publication probability. 

Besides the \textit{1-logit} and \textit{mlogit} selection models, \citet{begg1994operating} and~\citet{macaskill2001comparison} used the exponential selection models, 
\begin{equation*}
 p\left(y_i, s_i\right)=\exp \left\{-\beta \Phi^\gamma\left(-t_i\right)\right\}, 
\end{equation*}
with $\gamma$=1 or 2. We denote them by \textit{$\exp (\gamma=1)$} and \textit{$\exp (\gamma=2)$} selection models.

Compared with one-parameter selection models, two-parameter selection models have the advantage of describing the publication process that does not depend on the observed values~\citep{huang2023adjusting}. The two-parameter selection models include the two-parameter probit selection models
\begin{equation*}
 p\left(y_i, s_i\right)= \Phi(\alpha + \beta t_i)
\end{equation*}
and the two-parameter logistic selection model, 
\begin{equation*}
 p\left(y_i, s_i\right)=\frac{\exp \left(\alpha+\beta t_i\right)}{1+\exp \left(\alpha+\beta t_i\right)}.
\end{equation*}

We denote them by the \textit{2-probit} and \textit{2-logit} selection models, respectively. The above $t$-statistics type selection models are monotonic with the value of $t_i$, corresponding to the $p$-value of a one-sided test. In contrast, if one prefers a two-sided significance test, it is recommended to apply a modified selection model by changing $t_i$ into $|t_i|$. \citet{copas2013likelihood} proposed a likelihood-based sensitivity analysis with $t$-statistics type selection models. \citet{huang2023adjusting} proposed the inverse probability weighting method by the $t$-statistics type selection model using clinical trial registries. The validity of these methods relies on the correct specification of the parametric selection model. However, in practice, it is not easy to determine and select a definitely relevant selection model. Then, many researchers recommended applying multiple selection models in sensitivity analysis~\citep{huang2021using, huang2024publication, hu2024sensitivity}.

\subsection{Copas-Jackson non-parametric worst-case bound}
\label{sec:cj}

\citet{copas2004bound} first proposed a sensitivity analysis method studying the bounds for bias on the estimation for the overall effect size $\mu$ without assuming any parametric form of the selection models. Recall the marginal distribution of the random-effects model, $y_i \sim N(\mu, s_i^2 + \tau^2)$. Following the formulation by \citet{copas2004bound}, we denote $y \sim N(\mu, \sigma^2)$. We omit the subscript $i$ for the study index in the section following the original paper. Here $\sigma^2 =s^2+\tau^2$. Thus, $y\sim f(y\mid \sigma)=1/\sigma \cdot \phi\left\{ (y-\mu)/\sigma \right\}$, where $\phi(\cdot)$ means the probability density function (p.d.f.) of the standard normal distribution. Assuming that the population p.d.f. of $\sigma$ is $f(\sigma)$. The publication probability conditional on $y$ and $\sigma$ is denoted by $p(y, \sigma)=\mathrm{Pr}(\mathrm{selected}\mid y, \sigma)$; the selection probability conditional only on $\sigma$ is given by 
\begin{equation*}
p(\sigma)= \mathrm{Pr}(\mathrm{selected}\mid \sigma) = \int \mathrm{Pr}(\mathrm{selected}\mid y, \sigma) f(y\mid \sigma) dy =\int \frac{1}{\sigma} \phi\left(\frac{y-\mu}{\sigma}\right) p(y, \sigma) d y.
\end{equation*}
Recall that we call $p(y, \sigma)$ or $p(y, s)$ the selection model. To avoid confusion, we call $p(\sigma)$ the marginal representation of the selection model $p(y, \sigma)$, or simply the \textit{m-representation}. In the theory by \citet{copas2004bound}, $\tau$ is fixed and regarded as a sensitivity parameter. Then, $p(y, \sigma)$ can be denoted as $p(y, s)$ as done in Section~\ref{sec:sm}. By further marginalizing, the overall marginal selection probability is considered:
\begin{equation*}
 p = \mathrm{Pr}(\mathrm{selected}) = \int \mathrm{Pr}(\mathrm{selected}\mid \sigma)f(\sigma) d\sigma = \int p(\sigma)f(\sigma) d\sigma.
\end{equation*}
\citet{copas2004bound} assumed the following assumption.

\vspace{0.1cm}
\textbf{Assumption $A_0$:} $p(\sigma)$ is a non-increasing function of $\sigma$.
\vspace{0.1cm}

This assumption suggests studies with smaller $\sigma$ have a higher probability of being published and vice versa. 
\citet{copas2004bound} considered the class of the selection models satisfying Assumption $A_0$. In words, the class of the selection model consists of all the selection models $p(y, \sigma)$, whose \textit{m-representation} is monotone with respect to $\sigma$. Denote the class of the selection model by $\mathscr A_0$: 

\vspace{0.1cm}
$\boldsymbol{\mathscr A_0}$=$\{p(y, s)$: its \textit{m-representation} $p(\sigma)$ satisfies Assumption $A_0\}$.
\vspace{0.1cm}

Note that the C-J bound does not add any parametric assumption on the selection model. The lower $\sigma$ is often associated with larger studies. The assumption is based on empirical observations that larger studies are more likely to be published. In practice, $\mu$ is often estimated by maximizing the observed likelihood (ML) with all published studies. Denote the bias of the ML in the presence of selective publication as $b$. The bias is given by~\citet{copas2004bound} is 
\begin{equation}
 b=\frac{\int_0^{+\infty} \int_{-\infty}^{+\infty} \sigma^{-1} z \phi(z) p(\mu+\sigma z, \sigma) f(\sigma) d z d \sigma}{\int_0^{+\infty} \sigma^{-2} p(\sigma) f(\sigma) d \sigma},
 \label{eq.1}
\end{equation}
where $z=(y-\mu)/\sigma$ is introduced as the standard normal variable. \citet{copas2004bound} derived the analytic upper and lower bounds for the bias $b$ over the class of the selection models $\mathscr A_0$ under a given marginal selection probability $p$, that is, 
\begin{equation}
 |b| \leq \frac{\int_0^{+\infty} \sigma^{-1} p(\sigma) f(\sigma)}{\int_0^{+\infty} \sigma^{-2} p(\sigma) f(\sigma)} \frac{\phi\left\{\Phi^{-1}(p)\right\}}{p}=\frac{E_O\left(\sigma^{-1}\right)}{E_O\left(\sigma^{-2}\right)} \frac{\phi\left\{\Phi^{-1}(p)\right\}}{p}, \label{eq.2}
\end{equation}
where $E_O$ denotes the expectation conditional on the published. Assuming that there are $M$ unpublished studies, which leads to $p=N/(N+M)$. The sensitivity analysis can be interpreted as a caveat on how the estimates of effect size would be influenced by an increasing number of unpublished studies $M$. In their paper, the C-J bound, including the lower bound $L_{CJ}$ and upper bound $U_{CJ}$ can be estimated by empirical data, which is
\begin{equation}
 L_{CJ}=-\frac{\sum_{i=1}^N \sigma_i^{-1}}{\sum_{i=1}^N \sigma_i^{-2}} \frac{\phi\left\{\Phi^{-1}(p)\right\}}{p},~~U_{CJ}=\frac{\sum_{i=1}^N \sigma_i^{-1}}{\sum_{i=1}^N \sigma_i^{-2}} \frac{\phi\left\{\Phi^{-1}(p)\right\}}{p}, 
 \label{eq.3}
\end{equation}
where $\sigma_i=(s_i^2+\tau^2)^{1/2}$. In their proposal, $\tau$ is fixed as some plausible values and regarded as a sensitivity parameter. A choice of $\tau$ is the ML estimate without accounting for selective publication. 

\section{The connections between the assumption of the Copas-Jackson nonparametric worst-case bounds and existing selection models: does the C-J bound sufficiently work?}
\label{sec:connection}

Sensitivity analysis based on selection models plays an important role in revealing how much selective publication would affect the estimation of effect size and then result in misleading findings. The sensitivity analysis based on the C-J bound is useful and attractive since it requires minimal assumptions without any parametric distributional family for the selection models, disclosing the possible worst-case bias under its assumptions. However, we would wonder whether the assumption of the monotonicity of $p(\sigma)$ is satisfied by existing widely used selection models, or whether the class $\mathscr A_0$ covers a sufficiently wide class of selection models. We answer these two questions through both theoretical analysis and simulation studies with two popular selection models in sensitivity analysis, the Copas-Heckman selection model by~\citet{copas2000meta, copas2001sensitivity} and the \textit{2-probit} selection model by~\citet{copas2013likelihood}. 

We begin with the theoretical analysis to show if the expressions of the \textit{2-probit} and the Copas-Heckman selection model fulfill Assumption $A_0$. The \textit{m-representation} of the Copas-Heckman selection model is given by 
\begin{equation*}
 p(\sigma)=\Phi\left(\gamma_0+\frac{\gamma_1}{s}\right)~~\forall \tau.
\end{equation*}
Here, we assume $\gamma_1\geq 0$. Thus, studies with smaller reported variances and larger sample sizes are more likely to be published. As the $\sigma=(s^2+\tau^2)^{1/2}$ is an increasing, one-to-one correspondence function of $s~(s>0)$ given $\tau^2$ fixed, and $p(\sigma)$ is a non-increasing function of $s$ under $\gamma_1\geq 0$; it follows that $p(\sigma)$ is also a non-increasing function of $\sigma$. Therefore, the Copas-Heckman selection model~\eqref{eq.heckman} satisfies Assumption $A_0$ or belongs to the class $\mathscr A_0$ .

For the \textit{2-probit} selection model, the \textit{m-representation} is given by
\begin{equation*}
 p(\sigma)=\Phi\left\{\frac{\alpha+{\beta \mu}/{s}}{\sqrt{1+\beta^2\left(1+{\tau^2}/{s^2}\right)}}\right\}.
\end{equation*}
It has a turning point at $s=\frac{\alpha \tau^2}{\beta\left(1+\beta^2\right) \mu}$ if $\frac{\alpha \tau^2}{\beta\left(1+\beta^2\right) \mu}>0$, and $p(\sigma)$ changes from increasing to decreasing at this point. Thus, $p(\sigma)$ may not be a non-increasing function of $\sigma$.
In Section~\ref{sec:simu}, we use simulation-based analysis to show that the C-J bound is inappropriate for the \textit{2-probit} selection model, which is out of Assumption $A_0$. 

\section{Bound for PB over a class of $t$-statistics type selection models}
\label{sec:bound}

\subsection{A new class of selection models}

We showed that the assumption $A_0$ is not necessarily satisfied by some representative parametric selection models. Since selection models monotone with respect to $t$-statistics are a very natural class of selective publication mechanism, it is attractive to evaluate the worst-case bounds over a class of the selection models monotone with respect to $t$-statistics. Denote the class of selection models monotone with respect to $t$-statistics by

\vspace{0.1cm}
$\mathscr B_1$=$\{p(y, s): p(y, s)$ is monotone with respect to $t=y/s\}$.

$\mathscr B_2$=$\{p(y, s): p(y, s)$ is monotone with respect to $|t|=|y/s|\}$.
\vspace{0.1cm}

It would be useful to extend the Copas-Jackson-type bounds over $\mathscr B_1$ or $\mathscr B_2$. However, it seems very difficult to obtain an analytical expression of the bounds over $\mathscr B_1$ or $\mathscr B_2$. Instead, we consider a new class of selection models, which contains all selection models monotone with respect to $t$-statistics, and a method to calculate the worst-case bounds over the new class. 

We begin with introducing a new concept on the selection model. Recall that the random-effects model is described as $y_i\sim N(\mu_i, s_i^2)$ and $\mu_i\sim N(\mu, \tau^2)$ in Section~\ref{sec:re}. Being consistent with the notation and framework in Section~\ref{sec:cj}, we represent the random-effects model as $y\sim N(\tilde{\mu}, s^2)$ and $\tilde{\mu}\sim N(\mu, \tau^2)$. 
For the selection model $p(y, s)$, we introduce the \textit{$\mu$-representation} of the selection model by 
\begin{equation}
 p_1\left(\tilde{\mu}, s\right)=\int p_0\left(y, s\right) \frac{1}{s} \phi\left(\frac{y-\tilde{\mu}}{s}\right) d y,
 \label{p1}
\end{equation}
where we denote the function $p(y, s)$ as $p_0(y, s)$ to be consistent with following notations, despite abuse of notations. Furthermore we denote the publication probability conditional only on $s$ by $p_2(s)$, 
\begin{equation}
 p_2\left(s\right)=\int p_1\left(\tilde{\mu}, s\right) \frac{1}{\tau} \phi\left(\frac{\tilde{\mu}-\mu}{\tau}\right) d \tilde{\mu}.
 \label{p2}
\end{equation}
We build the assumption on the \textit{$\mu$-representation} $p_1(\tilde{\mu}, s)$;



\vspace{0.1cm}
\textbf{Assumption $A_1$:}
There exists $\mu_*$ such that either of the following conditions holds: 
1. $p_1(\tilde{\mu}, s)$ is a non-increasing function of $s$ for $\tilde{\mu} \le \mu_*$ and non-decreasing for $\tilde{\mu} > \mu_*$. 2. $p_1(\tilde{\mu}, s)$ is a non-decreasing function of $s$ for $\tilde{\mu} \le \mu_*$ and non-increasing for $\tilde{\mu} > \mu_*$.
\vspace{0.1cm}

Define a new class of selection models:

\vspace{0.1cm}
$\boldsymbol{\mathscr A_1}$=$\{p(y, s)$: its \textit{$\mu$-representation} $p(\tilde{\mu}, s)$ satisfies Assumption $A_1\}$.
\vspace{0.1cm}

We prove that all $t$-statistics type selection models satisfie Assumption $A_1$, that is, $\mathscr B_1 \bigcup \mathscr B_2 \subset \mathscr A_1$ (see Web Appendix A). Thus, obtaining the maximum and the minimum of $b$ over $\mathscr A_1$ is a relaxation problem for the original problem. They provide upper and lower bounds for the bounds over the class of selection models monotone with respect to $t$-statistics although they may not be tight. We found that only the Copas-Heckman selection model may violate the Assumption $A_1$ (see Appendix A) among the models in Section~\ref{sec:pre}. It is covered by the framework of~\citet{copas2004bound} with Assumption $A_0$. Thus, we construct the general class of selection models by $\mathscr A_0 \bigcup \mathscr A_1$ which covers all the selection models listed in Section~\ref{sec:pre}. 

\subsection{Bounds for bias under the new assumption}

It is difficult to derive an explicit expression of the worst-case bounds for PB over the class $\mathscr A_1$, thus, we propose an algorithm instead. Given different marginal selection probabilities, $p$, there exists a connection between the selection models and $p$ as follows:
\begin{equation}
 p=\int p_2\left(s\right) f\left(s\right) d s, 
 \label{eq.4}
\end{equation}
where $f(s)$ is the unknown distribution of the within-study standard error $s~(s>0)$. The bias is given by
\begin{equation}
b=\frac{\int_0^{+\infty} {1}/{\left(s^2+\tau^2\right)} \iint \left(s z+\tau w\right) p_0\left(s z+\tau w+\mu, s\right) \phi\left(z\right) \phi\left(w\right) d z d w f(s) d s}{\int_0^{+\infty} {1}/{\left(s^2+\tau^2\right)} p_2(s) f(s) d s}
 \label{eq.5}
\end{equation}
where $z =(y - \tilde{\mu})/s$ and $w=(\tilde{\mu}-\mu)/\tau$ are two standard normal variables independent with each other. The detailed derivation is presented in Web Appendix B. We use $s$ to denote the random variable for the within-study standard error, where $s_i$ is its sample. Let $f_o(s)$ denote the density function for $s$ conditional on the published. It has the following association with the overall selection probability:
\begin{equation}
 f_o(s)=\frac{\operatorname{Pr}(\text { select } \mid s) f(s)}{\operatorname{Pr}(\text { select })}=\frac{p_2(s) f(s)}{p},
 \label{eq.6}
\end{equation}
Thus, the expression of the bias~\eqref{eq.5} translates to 
\begin{equation}
 b=\frac{\int_0^{+\infty} {1}/{(s^2+\tau^2)}\cdot {f_o(s)}/{p_2(s)} \left\{\iint \left(s z+\tau w\right) p_0\left(s z+\tau w+\mu, s\right) \phi\left(z\right) \phi\left(w\right) d z d w\right\} d s}{\int_0^{+\infty} {1}/{\left(s^2+\tau^2\right)} f_o(s) d s}.
 \label{eq.7}
\end{equation}

Due to the complexity of~\eqref{eq.7}, it seems difficult to obtain an analytic bound for $b$ as done by~\citet{copas2004bound}. 
We can use a simulation-based method to derive the bounds of the bias as proposed by~\citet{zhou2024nonparametric}. Recall that $s_i$'s $(i=1, 2, \cdots, N)$ are the standard error of $y_i$ from the published studies, and they are random samples from the distribution $f_o(s)$. Thus, the theoretical expectation with respect to $f_o(s)$ in~\eqref{eq.7} can be approximated by the sample mean over $\{s_i\}$. We generate $K_1$ and $K_2$ samples for $w$ and $z$ independently from the standard normal distributions, which are indexed by $w_{k_1} (k_1=1, \cdots, K_1)$ and $z_{k_2} (k_2=1, \cdots, K_2)$, respectively. 
Denote $p_{i, k_1, k_2}=p_0(\mu+\tau w_{k_1}+s_i z_{k_2}, s_i)$, $p_{i, k_1}=p_1(\mu+\tau w_{k_1}, s_i)$, and $p_i=p_2 (s_i)$. 
Recall that quantities with $i=1,2, \cdots, N$ were for published studies and those with $i=N+1, \cdots, N+M$ are for unpublished studies. Then, we have an approximate expression of the bias:
\begin{equation}
 b=\frac{\sum_{i=1}^N {1}/{(s_i^2+\tau^2)}\cdot {1}/{p_i} \sum_{k_1=1}^{K_1} \sum_{k_2=1}^{K_2}\left(s_i z_{k_2}+\tau w_{k_1}\right) p_{i, k_1, k_2}}{\sum_{i=1}^N {1}/{(s_i^2+\tau^2)}}.
 \label{eq.8}
\end{equation}

By changing variables $\tilde{\mu}=\mu+\tau w$ and $y = \tilde{\mu} + s z$, \eqref{p1} and~\eqref{p2} lead to 
\begin{equation}
 p_1\left(\mu+\tau w, s\right)=\int p_0\left(\mu+\tau w+s z, s\right) \phi\left(z\right) d z,
 \label{p1b}
\end{equation}
and
\begin{equation}
 p_2\left(s\right)=\int p_1\left(\mu+\tau w, s\right) \phi\left(w\right) d w.
 \label{p2b}
\end{equation}
Then there are the following associations among these variables. $p_{i, k_1}=\int p_0(\mu+\tau w_{k_1}+s_i z, s_i )\phi(z)dz = 1/K_2 \sum_{k_2=1}^{K_2} p_0(\mu+\tau w_{k_1}+s_i z_{k_2}, s_i )=1/K_2 \sum_{k_2=1}^{K_2} p_{i, k_1, k_2}$, and by similar derivations, it holds $p_i=1/K_1 \sum_{k_1=1}^{K_1} p_{i, k_1}$
Given the marginal selection probability $p$, we may consider the following optimization problem;
$$
\texttt{OPT1}: \max /\min~b
$$
subject to\\ 
(C1) $0 \leq p_{i, k_1, k_2} \leq 1$; \\
(C2) $p_{i, k_1}=1 / K_2 \sum_{k_2=1}^{K_2} p_{i, k_1, k_2} ; p_i=1 / K_1 \sum_{k_1=1}^{K_1} p_{i, k_1}$;\\
(C3) $\exists k^{\prime} \in\left\{1, \cdots, K_1\right\}$, s.t. $p_{i, k_1} \leq p_{j, k_1}$ for $s_i \geq s_{j}$ when $z_k \leq z_{k^{\prime}}$, and $p_{i, k_1} \geq p_{j, k_1}$ for $s_i \geq s_{j}$ when $z_k \geq z_{k^{\prime}}$, or $p_{i, k_1} \geq p_{j, k_1}$ for $s_i \geq s_{j}$ when $z_k \leq z_{k^{\prime}}$, and $p_{i, k_1} \leq p_{j, k_1}$ for $s_i \geq s_{j}$ when $z_k \geq z_{k^{\prime}}$;\\
(C4) $p = 1 / (N+M) \sum_{i=1}^{N+M} p_i$.

The condition (C1) implies the property of probabilities, and (C2) is from~\eqref{p1b} and~\eqref{p2b}. Condition (C3) holds because of the Assumption $A_1$. Condition (C4) comes from~\eqref{eq.4}. However, Condition (C4) is not tractable since $p_i$ is not available for unpublished studies ($i=N+1, \cdots, N+M$). Instead of \texttt{OPT1}, we consider the following optimization problem;
$$
\texttt{OPT2}: \max / \min ~ b
$$
subject to (C1), (C2), (C3) and
(C4$^{\prime}$) $p \leq 1 / N \sum_{i=1}^N p_i$. The last condition (C4$^{\prime}$) is from the fact
\begin{equation}
\begin{aligned}
 \frac{1}{N} \sum_{i=1}^N p_i=E_O\left\{p_2(s)\right\}&=\int p_2(s) f_o(s) d s=\frac{1}{p} \int\left\{p_2(s)\right\}^2 f(s) d s \\& \geq \frac{{1}/{p}\left[\left\{\int p_2(s) f(s) d s\right\}^2\right]}{\int f(s) d s}=\frac{1}{p} \frac{p^2}{1} =p
\end{aligned}
\end{equation}
where $E_O$ denotes the expectation over the published studies. The third equality holds due to~\eqref{eq.6}, and the inequality holds by Schwartz’s Inequality. It implies that \texttt{OPT2} is a relaxation problem of \texttt{OPT1}. Furthermore, proof of the equivalence of the two optimization problems is referred to Web Appendix A of~\citet{zhou2024nonparametric}. Thus, they give the same maximum and minimum values. 
In practice, we obtain the minimum and maximum of the simulation-based bias~\eqref{eq.8} given a fixed $p$ with the nonlinear optimization methods, which can be handled by software for nonlinear programming such as \texttt{OPTMODEL} procedure in SAS (version 9.4). We set $K_1=K_2=1000$ for examples in this paper. Let $U_{\mathscr A_1}$ denote the simulation-based lower bound through \texttt{OPT2} and $L_{\mathscr A_1}$ denote the lower bound. The upper and lower bound for bias over $\mathscr A_0 \bigcup \mathscr A_1$, which we call the extended bound, can be obtained by combining the simulation-based bounds and the C-J bound, given by 
\begin{equation}
  L_{\mathrm{extended}} = \min (L_{\mathscr A_1}, L_{CJ}), ~~ U_{\mathrm{extended}} = \max (U_{\mathscr A_1}, U_{CJ}).
\end{equation}


\section{Simulation studies}\label{sec:simu}

We conducted simulation studies to validate the effectiveness of the extended bound. We assumed the true selection model as \textit{2-probit} or Copas-Heckman. For the Copas-Heckman selection model, we set two sets of parameters $(\mu, \tau, \rho, \gamma_1)=(1.5, 2.5, 0.8, 0.2)$ and $(1, 1, 0.2, 2)$ to reflect the situations with small $\gamma_1$ and relatively large $\gamma_1$. The resulting experiments are named as Expe\_H\_1 and Expe\_H\_2, respectively. For the \textit{2-probit} selection model, we set $(\mu, \tau, \beta)=(1.5, 2.5, 4)$ and $(1, 1, 2)$ , and the resulting experiments are called the Expe\_P\_1 and Expe\_P\_2, respectively. For each experiment, we simulated 1000 independent meta-analyses consisting of $S=25$ published and unpublished studies. The within-study standard deviations were sampled from the log-normal distribution $LN(0, 0.5)$. We sampled the empirical effect sizes in each study with $y_i\sim N(\mu, \tau^2+s_i^2)$. The complete meta-analysis dataset without a selective publication is denoted by $\{y_i, s_i\}_{i=1}^S$. We consider selection functions which satisfy a given marginal selection probability $p$. To this end, we determine $\gamma_0$ in Expe\_H\_1 and Expe\_H\_2 and $\alpha$ in Expe\_P\_1 and Expe\_P\_2 by solving $p=1/S\sum_{i=1}^S p(y_i, s_i )$. We then decided if the study was published with a Bernoulli distributed variable $D_i\sim \operatorname{Bern}\left\{p(y_i, s_i)\right\}$. The published datasets are constituted by those with $D_i=1$, denoted by $\{y_i, s_i\}_{D_i=1}$ in each meta-analysis. We varied $p$ from 0.1 to 0.9 to form a series of published meta-analysis datasets.

We compared the performance of the C-J bound and the extended bound on these published meta-analysis datasets. We first estimated the parameters $(\mu, \tau)$ by maximizing the likelihood with only published studies. The bias of the maximum likelihood estimator due to selective publication was estimated by the difference between the estimated $\hat{\mu}$ with only published datasets and the estimate with both published and unpublished meta-analyses. We also estimated the C-J bound and the extended bound on each selected published meta-analysis dataset according to~\eqref{eq.3} and \texttt{OPT2}. In practice, the marginal selection probability $p$ is unknown and estimates under various $p$ would be evaluated as a sensitivity analysis. To evaluate the performance of the proposed method, we estimate the C-J bound and the extended bound in~\eqref{eq.3} and \texttt{OPT2} under the correctly specified $p$. For example, for the published studies selected under $p=0.1$, we evaluated the results with $p=0.1$ in~\eqref{eq.3} and \texttt{OPT2}. The parameter $\tau$ was adopted as the ML with only published studies. For each meta-analysis, we could obtain an absolute value of the estimate of bias, the C-J bound, and the extended bound; thus, 1000 values of absolute bias (the absolute value of the bias), the C-J bound, and the extended bound were available under each $p$ in each experiment. 

\begin{table}
 \centering
 \caption{The rate of bias-exceeding-the-bound (\%) among 1000 simulated meta-analyses with the C-J bound and the extended bound.}
 \label{tab:simu}

 \resizebox{0.9\linewidth}{!}{
 \begin{tabular}{ccccccccc}
\hline \multirow{2}{*}{$p$} & \multicolumn{2}{c}{Expe\_H\_1} & \multicolumn{2}{c}{Expe\_H\_2} & \multicolumn{2}{c}{Expe\_P\_1} & \multicolumn{2}{c}{Expe\_P\_2} \\
\cline{2-9} 
& C-J & Extended & C-J & Extended & C-J & Extended & C-J & Extended \\
\hline 0.9 & 6.3 & 0.7 & 1.2 & 0.0 & 62.9 & 9.6 & 49.1 & 0.8 \\
0.8 & 4.5 & 0.4 & 0.2 & 0.0 & 82.8 & 7.9 & 60.6 & 0.2 \\
0.7 & 3.7 & 0.1 & 0.5 & 0.0 & 91.2 & 7.5 & 60.8 & 0.2 \\
0.6 & 3.0 & 0.0 & 0.3 & 0.0 & 93.5 & 4.8 & 64.2 & 0.1 \\
0.5 & 3.7 & 0.0 & 1.0 & 0.0 & 93.8 & 1.9 & 69.3 & 0.1 \\
0.4 & 5.1 & 0.5 & 1.8 & 0.0 & 93.9 & 2.5 & 68.1 & 0.1 \\
0.3 & 8.5 & 0.4 & 2.8 & 0.1 & 92.1 & 4.4 & 67.7 & 0.6 \\
0.2 & 17.0 & 3.6 & 6.6 & 1.8 & 90.0 & 14.6 & 67.8 & 3.5 \\
0.1 & 31.2 & 18.6 & 18.2 & 7.4 & 87.7 & 49.4 & 71.6 & 25.6 \\
\hline
\end{tabular}
 }
\end{table}

To quantify the performance of the extended bound and the C-J bound on covering the bias, we summarized the rates of meta-analyses among the 1000 datasets of which the absolute bias has exceeded the C-J bound or the extended bound in Table~\ref{tab:simu}. We found that the C-J bound can cover most of the biases ($>90$\%) when $p\geq 0.3$ on Expe\_H\_1 and Expe\_H\_2 with Copas-Heckman selection models. An extreme $p$ may not be often seen in real-world meta-analysis and it is recommended to avoid an extreme $p$ in previous studies~\citep{copas2013likelihood}. Thus, the C-J bound covers the bias suggested by the Copas-Heckman selection models in most cases. In contrast, the rate of bias-exceeding-the-bound for the \textit{2-probit} selection models is relatively high for all $p$'s. The results indicate that the C-J bound in over 90\% of simulated datasets cannot cover the bias if $(\mu, \tau, \beta)=(1.5, 2.5, 4)$ and in over 60\% of simulated datasets cannot cover the bias if $(\mu, \tau, \beta)=(1, 1, 2)$. However, our extended bound obtained a satisfactory coverage rate among these simulated datasets. Only under an extremely low $p$, our extended bound may fail to cover the actual bias in more than 10\% simulated datasets. The results substantiated the effectiveness of the extended bound on \textit{2-probit} selection model which violate Assumption $A_0$ while belonging to the class $\mathscr A_1$.

Our extended bound is constructed over a wider class of selection models and as aforementioned, may not be a sharp bound for $t$-statistics type selection models. To quantify how large our extended bound is, we show the ratio of the length of the extended bound against that of the C-J bound, 
$$
r = \frac{ U_{\mathrm{extended}} - L_{\mathrm{extended}} }{U_{CJ} - L_{CJ}},
$$
and show the ratio under $p=0.1, \cdots, 0.9$ for each experiment in Table~\ref{tab:simu_ratio}. The results shows that the extended bound is of 1 to 3 times of the length of the C-J bound in most scenarios. 

\begin{table}
  \centering

  \caption{The summary ratio of the length of the extended bound against that of the C-J bound for 1000 simulations under $p=0.1, \cdots, 0.9$}
  \label{tab:simu_ratio}
  \begin{tabular}{ccccc}
\hline \multirow{2}{*}{$p$} & Expe\_H\_1 & Expe\_H\_2 & Expe\_P\_1 & Expe\_P\_2 \\
 & Median [Q1, Q3] & Median[Q1, Q3] & Median [Q1, Q3] & Median [Q1, Q3] \\
\hline 0.9 & 1.35[1.18,1.50] & 1.63[1.40,1.85] & 1.39[1.23,1.56] & 1.76[1.51,2.03] \\
0.8 & 1.33[1.24,1.42] & 1.81[1.60,2.12] & 1.41[1.31,1.52] & 2.16[1.84,2.69] \\
0.7 & 1.39[1.30,1.50] & 2.53[2.16,3.00] & 1.52[1.40,1.67] & 3.41[2.81,4.32] \\
0.6 & 1.79[1.61,2.03] & 2.96[2.58,3.37] & 2.13[1.84,2.45] & 3.83[3.24,4.75] \\
0.5 & 2.19[1.99,2.45] & 3.02[2.70,3.47] & 2.56[2.27,2.96] & 3.87[3.34,4.59] \\
0.4 & 2.41[2.18,2.73] & 2.99[2.69,3.43] & 2.72[2.44,3.14] & 3.62[3.16,4.33] \\
0.3 & 2.47[2.21,2.82] & 2.79[2.49,3.17] & 2.70[2.43,3.04] & 3.22[2.79,3.86] \\
0.2 & 2.31[2.04,2.67] & 2.50[2.20,2.79] & 2.41[2.18,2.72] & 2.73[2.30,3.19] \\
0.1 & 1.68[1.30,2.11] & 1.74[0.93,3.05] & 1.76[1.51,2.05] & 1.73[1.43,2.19] \\
\hline
\end{tabular}
\end{table}

Furthermore, we experimented on four additional settings with the Copas-Heckman and \textit{2-probit} selection model: for the Copas-Heckman selection model, Expe\_H\_3 with $(\mu, \tau, \rho, \gamma_1) = (1, 0.5, 0.2, 2)$ and Expe\_H\_4 with $(\mu, \tau, \rho, \gamma_1) = (1, 0.3, 0.2, 1)$; for the \textit{2-probit} selection model, Expe\_P\_3 with $(\mu, \tau, \beta) = (1, 0.5, 2)$ and Expe\_P\_4 with $(\mu, \tau, \beta) =(1, 0.3, 1)$. All experiments were under $p=0.7$. These four experiments have smaller between-study variance compared with the previous experiments. We placed the experiment results in Web Appendix C. In Expe\_H\_3 and Expe\_H\_4, both the extended bound and the C-J bound still performed well in covering over 90\% biases. In Expe\_P\_3 and Expe\_P\_4, the C-J bound performed much better than in Expe\_P\_1 and Expe\_P\_2 but in almost all cases the rate of bias-exceeding-the-bound is still larger than 10\%, suggesting that the C-J bound might not work for $t$-statistics type selection model even with small between-study variance.

\section{Examples}
\label{sec:ex}

We illustrate the applicability of our proposal with two real-world meta-analyses. The first example is the meta-analysis used in~\citep{copas2004bound} with 14 randomized clinical trials (RCT) studying the effect of the use of prophylactic corticosteroids on improving the survival of premature infants. The RCT compared the occurrence of deaths of infants between those given the treatment and those under control. The number of deaths and total cases were documented in Table 1 of~\citep{copas2004bound}. We reused it in our paper as Table~\ref{tab:eg1}. From the results of the lnOR, 13 of 14 studies showed negative lnOR, suggesting the effectiveness of the treatment. The overall lnOR obtained with the ML was -0.480 (95\% CI: [0.707, -0.244]) and the ML estimate of the between-study heterogeneity $\tau$ was 0, indicating that there was no significant difference among the study-specific effect sizes and a fixed effect model was enough for this meta-analysis. 

\begin{table}
 \centering
 \caption{Data of the prophylactic corticosteroids study.}
 \label{tab:eg1}
 \begin{tabular}{lrrrc} 
 \hline
Study & Treatment & Control & Log(OR) & Precision \\
\hline 1 & $3 / 64$ & $12 / 58$ & -1.55 & 1.57 \\
2 & $1 / 69$ & $5 / 61$ & -1.49 & 1.07 \\
3 & $4 / 81$ & $11 / 63$ & -1.33 & 1.71 \\
4 & $14 / 131$ & $20 / 137$ & -0.35 & 2.72 \\
5 & $9 / 40$ & $11 / 42$ & -0.19 & 1.98 \\
6 & $6 / 95$ & $9 / 94$ & -0.43 & 1.88 \\
7 & $7 / 121$ & $13 / 124$ & -0.61 & 2.11 \\
8 & $3 / 67$ & $7 / 59$ & -0.97 & 1.48 \\
9 & $1 / 71$ & $7 / 75$ & -1.64 & 1.10 \\
10 & $0 / 23$ & $1 / 22$ & -1.19 & 0.60 \\
11 & $5 / 49$ & $4 / 31$ & -0.28 & 1.47 \\
12 & $8 / 56$ & $10 / 71$ & 0.03 & 2.00 \\
13 & $32 / 371$ & $34 / 372$ & -0.06 & 3.90 \\
14 & $36 / 532$ & $60 / 538$ & -0.54 & 4.56\\
\hline
\end{tabular}
\end{table}

\begin{figure}
 \centering
 \includegraphics[width=0.6\linewidth]{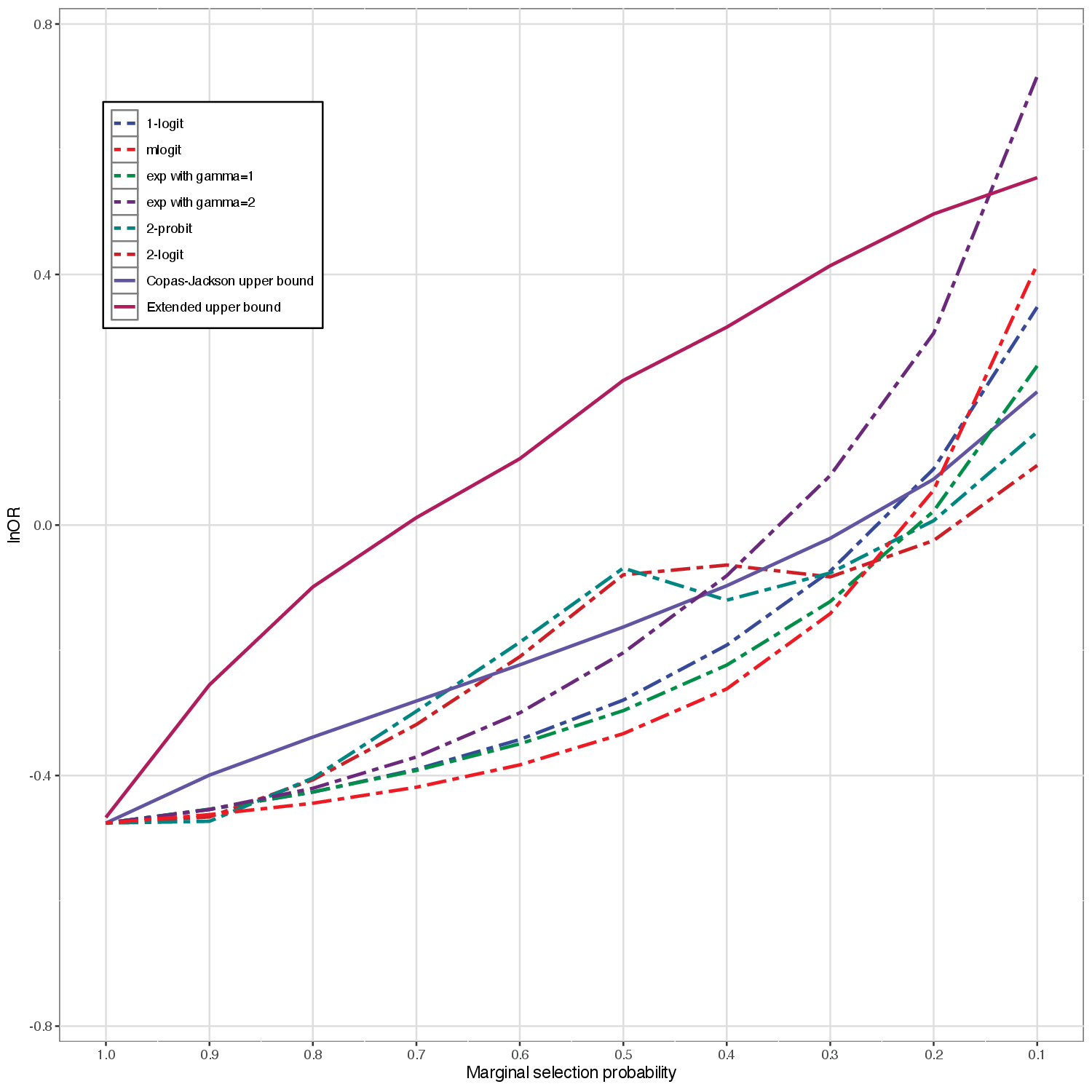}
 \caption{Variation of the estimates/bounds with the marginal selection probability $p$ for the meta-analysis of the prophylactic corticosteroids study (The relaxed upper bound is derived over $\mathscr A_0 \bigcup \mathscr A_1$, which is $\max (b_{CJ}, b_{\mathscr A_1})$).}
 \label{fig:eg1}
\end{figure}

We re-analyzed this example with our proposal under the relaxed assumptions. We gave the marginal selection probabilities $p=0.1, 0.2, \cdots, 0.9, 1$. Under each given $p$, we estimated the upper and lower bound of the bias based on both our proposed simulation-based bounds and the C-J bound. As we were more interested in the bounds of the estimates rather than the bounds for the bias, we thus gave the bounds of the estimates directly by adding the bounds of the bias on the estimates of $\mu=0.480$ without considering PB. Our proposal is an extension of the C-J bound to be applicable in more cases under a variety of widely used selection models. Therefore, we also conducted the sensitivity analysis on some widely used selection models including the one-parameter selection models \textit{1-logit}, \textit{mlogit}, \textit{$\exp \left(\gamma = 1\right)$} and \textit{$\exp \left(\gamma = 2\right)$}, and two two-parameter selection models \textit{2-probit} and \textit{2-logit}. We did not apply the Copas-Heckman selection models here as they relied on multiple sensitivity parameters that would make the comparisons difficult. Another reason for this was that the previous simulation studies showed that the bias suggested by the Copas-Heckman selection model was strongly covered by the C-J bound. We display the variation of each estimate and bound with $p$ decreasing in Figure~\ref{fig:eg1}. It has been shown that all the estimates of $\mu$ increase with $p$ decreasing so we only included the upper bound in Figure~\ref{fig:eg1}. None of the curves of estimated $\mu$ given by the above selection models was covered by the curve of the C-J bound, suggesting the C-J bound was insufficient in uncovering the maximal bias suggested by these selection models. Especially, for the exponential selection model with $\gamma=2$, the estimated $\mu$ began to exceed the C-J upper bounds when $p$ fell below 0.4. In contrast, the extended bound given by our proposal could almost cover all estimated $\mu$ from $p=0.9$ to $p=0.1$. Our extended bound only failed to cover the estimates given by the \textit{$\exp \left(\gamma = 2\right)$} selection models when $p\leq 0.1$. It validated the performance of our proposed extended bound. In addition, it suggested that even for studies with small between-study variance (in this example, $\tau^2$=0), the C-J bound is hard to cover the bias under all given $p$ if the true selection model is a $t$-statistics type selection model, verifying our concerns in the simulation studies. In contrast, our extended bound produces many acceptable worst-case bounds.

\begin{figure}
 \centering
 \includegraphics[width=0.6\linewidth]{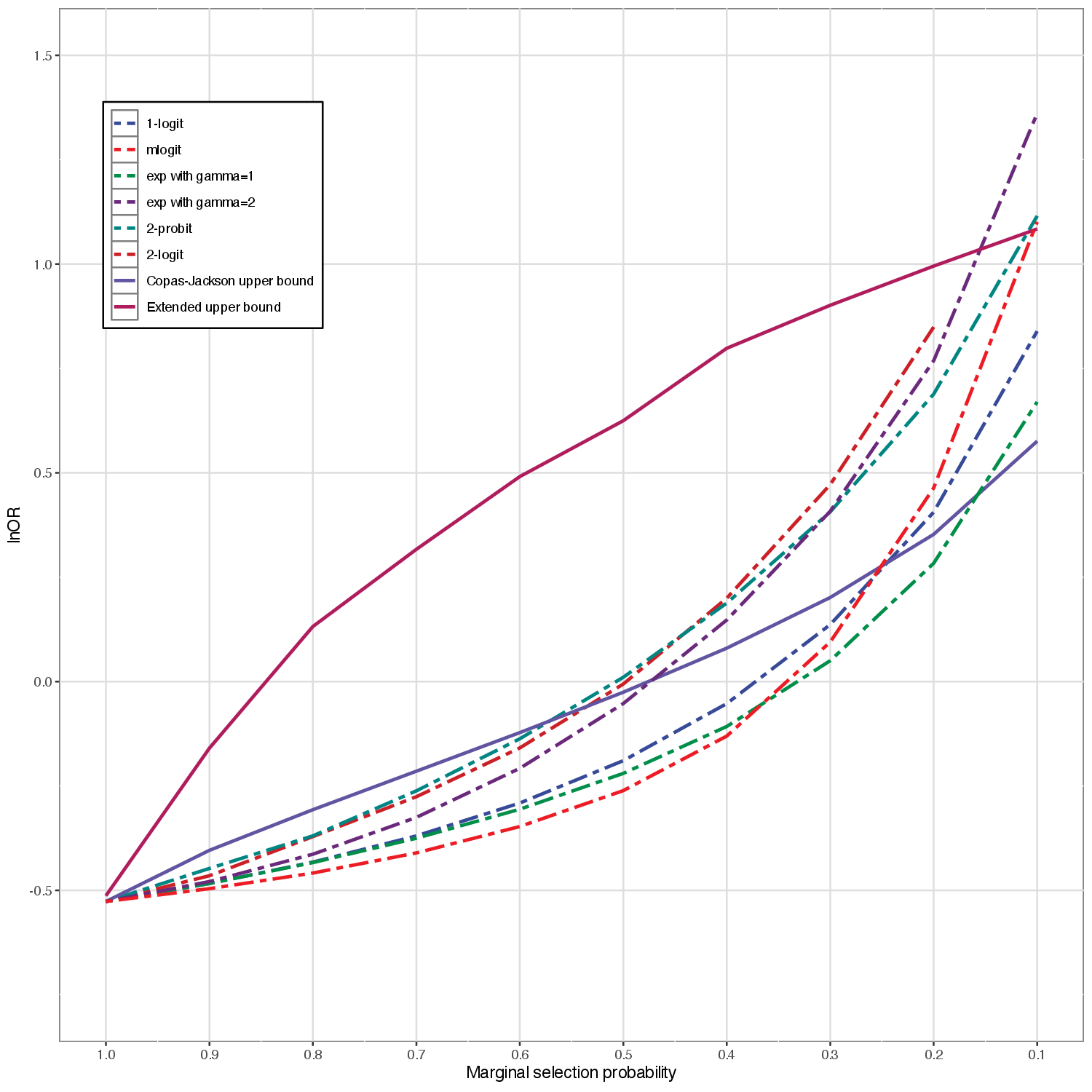}
 \caption{Variation of the estimates/bounds with the marginal selection probability $p$ on the meta-analysis for the Clopidogrel study (The relaxed upper bound is derived over $\mathscr A_0 \bigcup \mathscr A_1$, which is $\max (b_{CJ}, b_{\mathscr A_1})$).}
 \label{fig:clopi}
\end{figure}

We use another meta-analysis to examine the usefulness of our proposal in meta-analysis in the presence of small between-study heterogeneity. The meta-analysis consisting of 12 studies and conducted by~\citet{chen2013high} aimed to evaluate the treatment effect of the high-maintenance-dose Clopidogrel on major adverse cardiovascular/cerebrovascular events (MACE) compared with the standard dose. We presented the detailed data in Table~\ref{tab:clopi}. The overall effect size $\mu$ was estimated as -0.527 (95\% CI: [-0.930, -0.123]), and $\tau$ was estimated as 0.241. We conducted the sensitivity analysis with $p=0.1, 0.2, \cdots, 0.9, 1$ for both the selected selection models and the C-J bound and our extended bound over $\mathscr A_0 \bigcup \mathscr A_1$. The steps are similar to those for the previous example. We showed
the variation of PB-adjusted estimates/bounds in Figure~\ref{fig:clopi}. In this example, the C-J bound still could not cover the estimates given by the above selection models but ours could almost cover all PB-adjusted estimates for $p>0.1$, and meanwhile, an extremely low marginal selection probability is not realistic. The results suggested the applicability of our extended bound.

\begin{table}
 \centering
 \caption{Data of the Clopidogrel study.}
 \label{tab:clopi}
 \begin{tabular}{ccccccc}
\hline \multirow{2}{*}{ No. } & \multirow{2}{*}{ Study } & \multicolumn{2}{c}{ High dose } & \multicolumn{2}{c}{ Standard dose } & \multirow{2}{*}{Total subjects} \\
\cline { 3 - 6 } & & Events & Subjects & Events & Subjects & \\
\hline
 1 & Angiolillo 2008 & 0 & 20 & 0 & 20 & 40 \\
 2 & Aradi 2012 & 1 & 36 & 0 & 38 & 74 \\
 3 & RMYDA-150mg 201 & 0 & 25 & 0 & 25 & 50 \\
 4 & DOUBLE 2010 & 1 & 24 & 0 & 24 & 48 \\
 5 & EFFICIENT 2011 & 4 & 47 & 1 & 47 & 94 \\
 6 & GRAVITAS 2011 & 133 & 1109 & 113 & 1105 & 2214 \\
 7 & Han 2009 & 3 & 403 & 1 & 410 & 813 \\
 8 & Roghani 2011 & 3 & 205 & 2 & 195 & 400 \\
 9 & Tousek 2011 & 2 & 30 & 2 & 30 & 60 \\
 10 & VASP-02 2008 & 11 & 58 & 14 & 62 & 120 \\
 11 & von Beckerath 2007 & 2 & 31 & 2 & 29 & 60 \\
 12 & Wang 2011 & 9 & 150 & 25 & 156 & 306 \\
 \hline
\end{tabular}
\end{table}

\section{Discussions}
\label{sec:disc}

Selection models are useful in addressing PB in meta-analysis. They are widely used in both weighted PB-adjusted estimation and sensitivity analysis methods. Most methods rely on some parametric assumptions of the selection models. \citet{copas2004bound} developed the first sensitivity analysis method, the C-J bound, to allow a nonparametric form of the selection models based on only the non-decreasing of $p(\sigma)$ and propose a handy implementation to estimate the bound. In this paper, we investigated whether the class of the selection models for the C-J bound covers the widely used selection models. We found the assumption of the C-J bound is not satisfied by at least the \textit{2-probit} selection model from the perspective of both the theoretical analysis and simulation studies. To address this issue, we propose a relaxed assumption to allow a general class of selection models and a simulation-based method to derive the bounds for the bias under the assumption. Our general class of selection models covers most of the typical selection models, including the $t$-statistics type selection models with one-sided or two-sided tests and the Copas-Heckman selection models. We show the extended bound is reliable in covering PB suggested by these selection models with two real-world examples. Our proposed method is an efficient way to inspect the maximal bias caused by selective publication and evaluate the robustness of the results given by meta-analysis with published studies. 


The novelty of this paper is to extend the assumption of~\citet{copas2004bound} to the relaxed assumption and proposing a simulation-based algorithm to derive the extended bound. We verified the class of selection models under our assumption can cover the selection models in Section~\ref{sec:sm}. The verification of other selection models is lacking in this paper, which is left in our future work. Under the original Assumption $A_0$, \citet{copas2004bound} proposed an explicit expression for the bounds of bias given the marginal selection probability and estimated it with the empirical estimator. Under the alternative assumption $A_1$, it is difficult to derive the explicit expression for bounds. To address this issue, we derive the simulation-based bounds. The simulation-based procedure follows the large number theory and will give us almost accurate estimates for the bounds when the numbers of random numbers $z,w$ are sufficiently large. To more accurately estimate the bias, a large $(K_1, K_2)$ is required. However, too large $(K_1, K_2)$ either breaks the program down by introducing an \texttt{Out of memory} error in SAS and or takes a long time to find the optimum. Thus, we use $K_1=K_2=1000$ in practice. 

Another issue that exists in both the original C-J bound and our method is the estimates of between-study heterogeneity $\tau$. Both methods apply the naïve method using an appropriate estimate $\hat{\tau}$. Our proposal and \citet{copas2004bound} use the ML estimate for $\hat{\tau}$ without considering PB. The method ignores the relevance between the estimate for $\tau$ and $\mu$; it also does not address the potential bias in the estimates for $\tau$ in the presence of selective publication. It is recommended to evaluate the effect of $\hat{\tau}$ on the estimates of the bounds for the bias of $\mu$ by applying some plausible values to $\hat{\tau}$ and conducting a second-level sensitivity analysis. It is more appealing to develop novel methods that directly incorporate the uncertainty of $\tau$ when obtaining the bounds of bias for $\mu$. 

Sensitivity analysis methods have been widely used to address PB in meta-analysis including the meta-analysis for intervention studies~\citep{copas2001sensitivity, copas2013likelihood, ning2017maximum, hu2024sensitivity}, multiple treatments comparison~\citep{chootrakool2011meta}, meta-analysis of diagnostic studies~\citep{hattori2018sensitivity, piao2019copas, zhou2023likelihood}, and the network meta-analysis~\citep{marks2022embrace, mavridis2013fully, mavridis2014selection}. These sensitivity analyses are based on parametric selection models. It is relatively easy to estimate the confidence intervals (CI) with parametric selection models through the maximum likelihood theory. The CIs reflect the uncertainty of the estimates and help researchers better interpret the results. \citet{henmi2007confidence} was the first study to propose an estimation for the CI on the C-J bound. It will be appealing to develop a novel CI estimation under our assumption in the future work.

\section*{Acknowledgements}
This research was partly supported by Grant-in-Aid for Challenging Exploratory Research (16K12403) and for Scientific Research (16H06299, 18H03208) from the Ministry of Education, Science, Sports and Technology of Japan.




\bibliographystyle{biom} 
\bibliography{biomtemplate}

\label{lastpage}

\end{document}